\documentclass[11pt]{article}
\usepackage{amssymb}
\usepackage{amsfonts}
\usepackage{graphicx}
\usepackage{amsmath}
\usepackage{hyperref}

\makeatletter \@addtoreset{equation}{section} \makeatother

\newtheorem{theorem}{Theorem}
\newtheorem{lemma}{Lemma}

\newtheorem{proposition}{Proposition}

\begin{document}

\title{Change of variables as a method to study general $\beta$-models: bulk universality}
\author{ M. Shcherbina,
Institute for Low Temperature Physics Ukr.Ac.Sci, Kharkov, Ukraine.\\
E-mail:shcherbi@ilt.kharkov.ua }

\date{}

\maketitle

\begin{abstract}
We consider  $\beta$ matrix models with  real analytic potentials. Assuming that  the corresponding
equilibrium density $\rho$ has a one-interval support
(without loss of generality $\sigma=[-2,2]$), we study the transformation of the correlation functions after the change of variables
 $\lambda_i\to\zeta(\lambda_i)$ with $\zeta(\lambda)$ chosen from
the equation $\zeta'(\lambda)\rho(\zeta(\lambda))=\rho_{sc}(\lambda)$, where $\rho_{sc}(\lambda)$ is the standard
semicircle density.  This gives us the "deformed" $\beta$-model which has an additional "interaction" term.
Standard transformation with the Gaussian integral  allows us to show that the "deformed" $\beta$-model may be reduced
 to the standard Gaussian $\beta$-model with  a small  perturbation  $n^{-1}h(\lambda)$.
This  reduces most of the problems of local and global regimes for  $\beta$-models to the corresponding problems
for the Gaussian $\beta$-model with a small perturbation.
In the present paper we prove the bulk universality of local eigenvalue statistics for both one-cut and multi-cut cases.
\end{abstract}

\section{Introduction and main results}\label{s:1}
For any $ \beta>0$ we consider the distribution in $ \mathbb{R}^n$ of the form
\begin{align}\label{p_n}
p_{n,\beta}(\bar\lambda)=Z_{n,\beta}^{-1}[V]e^{\beta H(\bar\lambda)/2},
\end{align}
where    $ H$ (Hamiltonian) and
 $ Z_{n}[\beta,V]$ (partition function) are
\begin{align}\label{H}
H(\bar\lambda)=& -n\sum_{i=1}^n
V(\lambda_i)+\sum_{i\not=j}\log|\lambda_i-\lambda_j|,
\\
V(\lambda)>&(1+\varepsilon)\log(1+\lambda^2),
\notag\\
\label{Z_n}
 Z_{n}[\beta,V]=& \int e^{\beta H(\bar\lambda)/2}d\bar\lambda.
\end{align}
For any integrable function $\Phi(\bar\lambda)$ we denote its expectation by
\begin{align}\label{la-ra}
\langle\Phi(\bar\lambda)\rangle_{V,n}=\int\Phi(\bar\lambda)p_{n,\beta}(\bar\lambda)d\bar\lambda.
\end{align}
The expectation is closely connected with the correlation functions
\begin{align}\label{p_nl} p^{(m)}_{n,\beta}(\lambda_1,...,\lambda_m)=
\int_{\mathbb{R}^{n-l}} p_{n,\beta}(\bar\lambda) d\lambda_{m+1}...d\lambda_n.
\end{align}
It will be convenient below to use also the notation
\begin{align}\label{N_n}
\mathcal{N}_n[h]=\sum_{i=1}^n h(\lambda_i)
\end{align}
for the linear eigenvalue statistics, corresponding to the test function $h$.

For  $ \beta=1,2,4$ (\ref{p_n})-(\ref{Z_n}) is a joint eigenvalue distribution  of real symmetric,
hermitian and symplectic matrix models
respectively.

Since the papers \cite{BPS:95,Jo:98} it is known that for any $\beta>0$
if $V$ is a H\"{o}lder function, then
\begin{equation}\label{BPS}n^{-2}\log Z_{n,\beta}[V]=\frac{\beta}{2}\mathcal{E}[V]+O(\log n/n),\end{equation}
where
\begin{equation}\label{E_V}
\mathcal{E}[V]=\max_{m\in\mathcal{M}_1}\bigg\{
L[dm,dm]-\int V(\lambda)m(d\lambda)\bigg\}=\mathcal{E}_V(m^*),
\end{equation}
and the maximizing measure $m^*$ (called the equilibrium measure) has a compact
support $\sigma:=\mathrm{supp\,}m^*$. Here and below
 we denote
\begin{align}\label{L[,]}
&L[\,dm,dm]=\int\log|\lambda-\mu|dm(\lambda) dm(\mu),\\
&L[f](\lambda)=\int\log|\lambda-\mu|f(\mu)d\mu, \quad L[f,g]=(L[f],g),
\notag\end{align}
where $(.,.)$ is a standard inner product in $L_2[\mathbb{R}]$. The support $\sigma$ can consist of one
interval (one-cut case) and many intervals (multi-cut case).
If $V'$ is a H\"{o}lder function, then the equilibrium measure $m^*$ has a density
$\rho$ (equilibrium density).  The support $\sigma$ and the density $\rho$ are uniquely defined by the conditions:
\begin{equation}\label{cond_rho}\begin{array}{l}\displaystyle
v(\lambda ):=2\int \log |\mu -\lambda |\rho (\mu )d\mu -V(\lambda )=\sup v(\lambda):=v^*,\quad\lambda\in\sigma,\\
v(\lambda )\le \sup v(\lambda),\quad \lambda\not\in\sigma,\hskip
2cm\sigma=\hbox{supp}\{\rho\}.
\end{array}\end{equation}
Without loss of generality we will assume below that $v^*=0$.

One of the most important questions of the theory of random matrices is the
universality conjecture for the local eigenvalue statistics. According to this conjecture, e.g., in the bulk of the spectrum,
 the behavior of the scaled correlation functions  (\ref{p_nl})
\begin{equation}\label{univ}
p^{(n)}_{k,\beta}(\lambda_0+x_1/(n\rho(\lambda_0)),...,\lambda_0+x_k/(n\rho(\lambda_0)))\end{equation}
in the limit $n\to \infty$ is universal, i.e., does not depend on $V$ and $\lambda_0$ and depends only on $\beta$.
The case $\beta=2$ is the simplest one, since
 for $\beta=2$  all correlation functions
of (\ref{p_nl}) can be expressed in terms of the reproducing kernel of the system
of polynomials orthogonal with a varying weight $e^{-n\beta V}$ (see e.g. \cite{Me:91}).
The orthogonal polynomial machinery, in particular, the Christoffel-Darboux formula and
Christoffel's function simplify considerably the studies of  marginal densities (\ref{p_nl}).
This allows to study for $\beta=2$ the local eigenvalue statistics in many different cases: bulk of the spectrum,
edges of the spectrum, special points, etc. (see \cite{PS:97},
\cite{PS:07},\cite{DKMVZ:99},\cite{BI:03},\cite{C-K:06},\cite{L-Lub:08}).

For $\beta=1,4$ the situation is more complicated. It was shown in \cite{Tr-Wi:98} that all
correlation functions can be expressed in terms of some $2\times 2$- matrix kernels. But the
representation is less convenient than that in the case $\beta=2$.
Therefore the universality
conjecture for $\beta=1,4$ was proven much later than for $\beta=2$.
There were a number of papers with improving results, first for monomials
$V=\lambda^{2m}+o(1)$,
(see  \cite{De-G:07}, \cite{De-G:07a},\cite{DGKV:07}), then for arbitrary real
analytic one-cut potentials (see \cite{S:08}, \cite{S:09})
and finally  for  multi-cut real analytic potentials (see \cite{S:11}).

Note, that  for $\beta=1,2,4$ it was shown that the convergence of the scaled correlation functions (\ref{univ})
is uniform in $(x_1,\dots, x_k)\in S$,
where $S$ is an arbitrary compact set in $\mathbb{R}^k$. There is also a more weak form of the universality, when the
limit $n\to\infty$ is taken after the integration  of the correlation function of (\ref{univ})  with a smooth compactly supported
function $\phi(x_1,\dots,x_k)$. To prove universality in this form, it suffices to
 consider the limits of the expectations of the functions of the form
\begin{align}\label{Phi_k}
&\Phi_k(\bar\lambda;\lambda_0)=\prod_{j=1}^k \Big(n^{-1}\sum_{i=1}^nn
\phi_j\big(n\rho(\lambda_0)(\lambda_i-\lambda_0)\big)\Big),
\,\,\lambda_0\in (-2+\varepsilon,2-\varepsilon),\end{align}
where $\phi_j(x)$ ($j=1,\dots k$) - are arbitrary smooth functions
with  compact supports.

In the series of recent papers \cite{BEY:11,EY:13,BEY:13} the bulk universality for  any $\beta>0$
in the case of one-cut potentials of the generic behavior, possessing 4 derivatives,
 was proven in the form (see \cite{BEY:13}, Theorem 2.5):
\[
\lim_{n\to\infty}(2n^{-1+\varepsilon})^{-1}\int_{-n^{-1+\varepsilon}}^{n^{-1+\varepsilon}}dt\langle\Phi_k(\bar\lambda;\lambda_0+t)\rangle_{V,n}=
\lim_{n\to\infty}\langle\Phi_k(\bar\lambda;\lambda_0)\rangle_{*,n},
\]
where $\varepsilon$ is an arbitrary small number, and here and below we denote $\langle\dots\rangle_{*,n}$
the expectation (\ref{la-ra}) for the Gaussian case $V^*(\lambda)=\frac{1}{2}\lambda^2$.
Recall that the generic behavior of the potential $V$ means that
its equilibrium density has the form
\begin{equation}\label{rho}
    \rho(\lambda)=\frac{1}{2\pi}P(\lambda)\Im X^{1/2}_\sigma(\lambda+i0),\quad
    \inf_{\lambda\in\sigma}|P(\lambda)|>0,\quad X_\sigma(z)= z^2-4,
\end{equation}
where we choose a branch of $X^{1/2}_\sigma(z)$ such that $X^{1/2}_\sigma(z)\sim z$, as $z\to+\infty$.
Moreover, the function $v$ defined by (\ref{cond_rho})
attains its maximum only if $\lambda $ belongs to  $\sigma $.

 We recall also   that for sufficiently smooth  $V$
the equilibrium density $\rho$ always
 has the form (\ref{rho}) (see, e.g., \cite{APS:01}).
 For real analytic $V$ the function $P$    is also real analytic
(\ref{rho})  and  can be represented in the form
\begin{equation}\label{P}
    P(z)=\frac{1}{2\pi i}\oint_\mathcal{L}\frac{V'(z)-V'(\zeta)}{(z-\zeta) X_\sigma^{1/2}(\zeta)}d\zeta.
    \end{equation}
Hence  generic behavior just means that $\rho$ has no zeros in the internal points of $\sigma$ and behaves like
square root near the edge points.

In the present paper we propose a different from \cite{BEY:11,EY:13,BEY:13} method,
 based on the analysis of the transformation of  (\ref{p_n}) under a
smooth change of variables $\lambda\to\zeta(\lambda)$. For a good choice of $\zeta$ (see (\ref{eq_z}))
we obtain that the partition function and all the correlation functions of (\ref{p_n}) can be expressed in terms
of the Hamiltonian
\begin{align}\label{new_H}
\tilde H^{(\zeta)}(\bar\lambda)=&H_n^*(\bar\lambda)+\Big(\frac{2}{\beta}-1\Big)\sum\log\zeta'(\lambda_j)
+\sum_{k=1}^\infty\eta_k\Big(\sum_j(\varphi_k(\lambda_j)-(\varphi_k,\rho_{sc}))\Big)^2,
\end{align}
where $H_n^*$ is the Hamiltonian of the form (\ref{p_n}), corresponding to $V^*(\lambda)=\lambda^2/2$,
and  $\{\eta_k\}_{k=1}^\infty$ and $\{\varphi_{k}(\lambda)\}_{k=1}^\infty$   are  eigenvalues
and eigenvectors of the integral operator in $L_2[\sigma_\varepsilon]$ ($\sigma_\varepsilon=[-2-\varepsilon,2+\varepsilon]$)
with the kernel
\begin{align}\label{L^z}
L^{(\zeta)}(\lambda,\mu):=\log\Big|\frac{\zeta(\lambda)-\zeta(\mu)}{\lambda-\mu}\Big|=
\sum_{k=1}^\infty\eta_k\varphi_k(\lambda)\varphi_k(\mu).
\end{align}
For sufficiently smooth  $\zeta(\lambda)$ the operator with this kernel is a compact operator with smooth eigenfunctions.
The rate of convergence $\eta_k\to 0,\,k\to\infty$, depends on the number of derivatives of $\zeta(\lambda)$, e.g. for
$\zeta(\lambda)\in C_l[\sigma_\varepsilon]$, we have $\eta_k=o( k^{-l-1/2}),\,k\to\infty$
(see \cite{GK:69} Chapter III, \S 10). Hence, restricting summation in (\ref{new_H}) by $M=M(n)$, we can provide
that the remainder is $o(n^{-2})$, and so it does not contribute to the correlation functions. Then, using the Gaussian
integration
formula (see (\ref{H-S})) for each $k=1,\dots M$, we can "linearize" the terms under the summation and obtain that
\begin{align}\label{int_Phi}
\left\langle\Phi_k(\bar\zeta)\right\rangle_{V,n}=&\Big(\frac{\beta}{8\pi}\Big)^{M/2}\int d\bar ue^{-\beta(\bar u,\bar u)/8}
\langle \Phi_k(\zeta(\bar\lambda))e^{\beta\mathcal{N}_n[\dot h_u]/2}\rangle_{*,n}\Big/I_n[\beta,\zeta],
\end{align}
where a "small perturbation" of $V_*$ $h_u(\lambda)$ (defined by (\ref{h_u})) depends linearly on the integration parameters
$\bar u$, and $I_n[\beta,\zeta]$ is the normalizing constant
\begin{align*}
I_n[\beta,\zeta]:=&\Big(\frac{\beta}{8\pi}\Big)^{M/2}
\int d\bar u e^{-\beta(\bar u,\bar u)/8} \langle e^{\beta\mathcal{N}_n[\dot h_u]/2}\rangle_{*,n};
\end{align*}

Note that a similar method was used in \cite{S:13} in the multi-cut case in order to linearize the term  which corresponds
to the "interaction" between different intervals of the spectrum.

The analysis of $\langle e^{\beta\mathcal{N}_n[h]/2}\rangle_{*,n}$ is based on the well-known result of \cite{Jo:98},
which we will use in the form, obtained in \cite{S:13}, Theorem 1.
\begin{theorem}\label{t:0}
Let $V$ be a real analytic one-cut potential of generic behavior with $\sigma=[-2,2]$,
 and $h$ satisfy one of the two conditions:

(i) $h$ is a real valued function with $ ||h'||_2,||h^{(6)}||_2\le  n^{1-\tilde\delta}$ (here and below
$||.||_2$ means the standard norm in $L_2[-2-\varepsilon/2,2+\varepsilon/2]$,
with some  small $\varepsilon>0$), $\tilde\delta>0$ is any small constant;

(ii) $h$ is  complex valued, $\quad(D_\sigma\Re h,\Re h)+(D_\sigma\Im h,\Im h)\le  c_*\log n\quad$
with some sufficiently small $c_*$,
and $|h'||_2,||h^{(6)}||_2\le \log^sn$ with some $s>0$.

Then we have
\begin{align}\label{Joh}
& \langle e^{\beta\mathcal{N}_n[\dot h]/2}\rangle_{V,n}=\exp\Big\{
(h,\nu_\beta)+\frac{\beta}{8}(\overline D_\sigma h,h)+n^{-\alpha}O\big(||h^{'}||_2^3\big)+
n^{-\alpha}O\big(||h^{(6)}||_2^3 \big)\Big\},\end{align}
where $\dot{h}:=h-(\rho,h)$, and $\alpha=1$ for the case (i) and $\alpha=1/2$ for the case (ii),
\begin{align}
&\overline D_\sigma=\frac{1}{2}(D_\sigma+D^*_\sigma),
\quad D_\sigma h(\lambda)=
\frac{X_\sigma^{-1/2}(\lambda)}{\pi^2 }\int_{\sigma}
\frac{h'(\mu)X_\sigma^{1/2}(\mu) d\mu}{(\lambda-\mu)},\label{bar_D}\end{align}
and $D^*_\sigma$ is the adjoint operator to $D_\sigma$ in $L_2(\sigma)$. We will use also the representation of
$\bar D_\sigma$ obtained in \cite{Jo:98}
\begin{align}
(\overline D_\sigma h,h)=\sum_{k=1}^\infty kh_k^2,\quad h_k=\frac{2}{\pi}\int_0^\pi h(2\cos\theta)\cos k\theta d\theta.
\label{bar_D.1}\end{align}
A non positive measure $\nu_\beta$ in (\ref{Joh}) has the form
\begin{align*}
& (h,\nu_\beta):=\big(1-\frac{\beta}{2}\big)\Big(\frac{1}{4}(h(-2)+h(2))-
\frac{1}{2\pi}\int_\sigma\frac{h(\lambda)d\lambda}{ \sqrt{4-\lambda^2}}-
\frac{1}{2}(D_\sigma\log P,h)\Big).
\end{align*}
It will be important in what follows that
$ \bar D_\sigma$ is a rank one perturbation of $-{L}_\sigma^{-1}$, where
$ {L}_\sigma$ is the integral operator defined by the kernel $ \log|\lambda-\mu|$ for the
interval $ \sigma$ (see \cite{S:13}):
\begin{align}
\label{DL}
{L}_{\sigma}\bar D_{\sigma} v=-v+\pi^{-1}(v,X^{-1/2}_{\sigma})
\mathbf{1}_{\sigma}.
%\notag
\end{align}
\end{theorem}
It is easy to understand  that in view of (\ref{int_Phi}) it suffices to prove
 that in  the domain which gives non vanishing
contribution in the integral (\ref{int_Phi}) we have
\begin{align}\label{**}
\Big| \frac{\langle \Phi_ke^{\beta\mathcal{N}_n[\dot h]/2}\rangle_{*,n}}{\langle e^{\beta\mathcal{N}_n[\dot h]/2}\rangle_{*,n}}-
 \langle \Phi_k\rangle_{*,n}\Big|\le
\tilde\varepsilon_n, \quad \tilde\varepsilon_n\to 0,\end{align}
or, by another words, that for the Gaussian potential $V^*$ the "small perturbation" $n^{-1}h_u$
does not change correlation functions. We prove (\ref{**}) in two steps. On the first step we  replace
$h_u(\lambda)$ by some linear function $c(h_u)\lambda$, and on the second use the result
of \cite{VV:09}, Theorem 1, which (after "translation" on the langauge of correlation functions)
states  that for  any sequence $\lambda_0^{(n)}$ such that $n^{2/3}|\,|\lambda_0^{(n)}|-2|\to\infty$, the integrated
correlation functions $\langle\Phi_k(\bar\lambda,\lambda_0^{(n)})\rangle_{*,n}$ of (\ref{Phi_k}) converges
to some universal limit, depending only on $\{\phi_j\}_{j=1}^k$. This limit  corresponds to the so-called $Sine_\beta$ process,
whose  definition is not important here (see \cite{VV:09} for the precise definitions and  results).
We will use two simple corollaries from the above statement:
\begin{align}\label{VV1}
&|\langle\Phi_k(\bar\lambda,\lambda_0)\rangle_{*,n}|\le C_\Phi,\\
&|\langle\Phi_k(\bar\lambda,\lambda_0+t/n)\rangle_{*,n}-\langle\Phi_k(\bar\lambda,\lambda_0)\rangle_{*,n}|\le\varepsilon_n\to 0,\quad n\to\infty,
\label{VV2}\end{align}
where the first bound is uniform for $\lambda_0\in[-2+\varepsilon,2-\varepsilon]$,
and the second relation is uniform in the same $\lambda_0$ and  $|t|\le n^{1-\delta}$ if
$\delta>0$ is fixed. Note, that (\ref{VV1}) and (\ref{VV2}) become evident if we assume the contrary
for some sequence of $\lambda_0^{(n)}$ and
obtain the contradiction with Theorem 1 of \cite{VV:09}.

The method briefly described above gives the following result
\begin{theorem}\label{t:1} Let $V$ be a real analytic one-cut potential with $\sigma=[-2,2]$ of generic behavior
and $\lambda_0\in [-2+\varepsilon,2-\varepsilon]$ with any small $\varepsilon>0$. Then  for any $k\ge 1$
and any  $\Phi_k(\bar\lambda,\lambda_0)$ of the form (\ref{Phi_k}) with smooth $\{\phi_j\}_{j=1}^k$
we have uniformly in
$\lambda_0\in [-2+\varepsilon,2-\varepsilon]$
\begin{align}\label{t1.1}
\lim_{n\to\infty}\langle\Phi_k(\bar\lambda,\lambda_0)\rangle_{V,n}=
\lim_{n\to\infty}\langle\Phi_k(\bar\lambda,0)\rangle_{*,n}.
\end{align}
\end{theorem}

\textbf{Remarks}:

(i) The method  of Theorem \ref{t:1} can be generalized to the case of non-analytic $V$,
for  which   $P$ of (\ref{rho}) possesses 5 derivatives. The reasons to prove here Theorem \ref{t:1} for
real analytic $V$ is that in this case $|\eta_k|\sim e^{-kc}$, thus we can take $M=[\log^2n]$ and do not care
about $M^p$ and the number of derivatives in the formulas. This allows to simplify the proof of Lemmas \ref{l:conv}
and \ref{l:re_h}. Moreover, for  $P$
with 5 derivatives, the result (\ref{Joh}) cannot be applied and we need to prove  a new form of (\ref{Joh}), which requires only
$2+\varepsilon$ derivatives of $h$ (in the sense of Sobolev spaces), but  gives the bound only of the order $n^{-\kappa}$
with small $\kappa>0$, instead of $n^{-1}$ in (\ref{Joh}). All these technicalities make the proof less straightforward.
Thus, the proof of Theorem \ref{t:1} for non analytic potentials is postponed to the next paper,
where the edge universality will be proved by the
same method.

(ii) Examining the proof of Theorem \ref{t:1}, it is easy to see that to prove the edge universality
for $\lambda_0=2$, it suffices just to
replace $\Phi_k$ of (\ref{Phi_k}) by
\[\tilde\Phi_k(\bar\lambda,)=\prod_{j=1}^k \Big(n^{-1}\sum_{i=1}^nn
\varphi_j\big(n^{2/3}\gamma_P(\lambda_i-2)\big)\Big),
\,\,\gamma_P=P^{2/3}(2)\]
The only difference will be that instead of (\ref{VV1}), (\ref{VV2}) we need to use similar results of \cite{RRV:07} on
the existence of  limits for the scaled correlation functions near the edge point for the Gaussian potential $V^*$.

(iii) The method proposed here  works  well for the global regime problems, e.g., for the proof of CLT
for $\mathcal{N}_n[h]$,  the expansion for $Z_n[\beta,V]$ in $n^{-k}$, computations of the Hankel and the Toeplitz
determinants etc. In particular, it simplifies considerably the proof of well-known results, (see, e.g.,
\cite{Jo:98,S:08,S:09,S:11,BG:11}) because it reduces their proofs to the case of the Gaussian potential $V^*$
with a small perturbations $\frac{1}{n}h$.

\smallskip

Using the results of \cite{S:13},  Theorem \ref{t:1} can be generalized to the multi-cut real analytic potentials
of generic behavior.
\begin{theorem}\label{t:2}  Let $V$ be a real analytic multi-cut potential with $\sigma=\cup_{\alpha=1}^q\sigma_\alpha$
($\sigma_\alpha=[a_\alpha,b_\alpha]$)
of generic behavior, which means that the correspondent equilibrium density $\rho$ has the form (\ref{rho}),
with $X_\sigma=\prod_{\alpha=1}^q(z-a_\alpha)(z-b_\alpha)$. Then, for any
 $\lambda_0\in \cup_{\alpha=1}^q[-a_\alpha+\varepsilon,b_\alpha-\varepsilon]$  (\ref{t1.1}) holds.

\end{theorem}
The paper is organized as follows. The proof of Theorem \ref{t:1} modulo few auxiliary statements (see Lemmas \ref{l:zeta}-\ref{l:3})
is given in Section 2. The proofs of Lemmas \ref{l:zeta}-\ref{l:3} are given in Section 3. The proof of Theorem \ref{t:2} is given
in Section 4.

\section{Proof of Theorem \ref{t:1}}
 Take any $n$-independent  small $\varepsilon>0$. It is known (see  \cite{BPS:95}) that if we
replace in the definition of the partition function and of the correlation functions
 the integration over $\mathbb{R}$
by the integration $\sigma_{\varepsilon/2}$, $p^{(m)}_{n,\beta}$ and the new  marginal
densities $p^{(m,\varepsilon)}_{n,\beta}$ for $m=1,2,\dots$
 satisfy the inequalities
\begin{align*}\notag
\sup_{\lambda_1,\dots,\lambda_m\in\sigma_{\varepsilon/2}}&|p^{(m)}_{n,\beta}(\lambda_1,\dots,\lambda_m)
-p^{(m,\varepsilon)}_{k,\beta}(\lambda_1,\dots,\lambda_m)|
\le C_me^{-n\beta d_\varepsilon},\label{t2.*}\\ & {Z_{n}[\beta, V]}={Z_{n}^{(\varepsilon)}[\beta,V]}
\big(1+e^{-n\beta d_\varepsilon}\big).
\notag\end{align*}
 It is more convenient  to consider the
integration with respect to $\sigma_{\varepsilon/2}$,
 thus, starting from this moment
 it is assumed  that this
truncation is made, and below the integration without limits means the
integration over $\sigma_{\varepsilon/2}$, but  the superindex $\varepsilon$ will be omitted.

\subsection{Change of variables in the one cut case }
Let $V$ be some smooth enough  potential with the equilibrium density $\rho$, $\hbox{supp}\rho=[-2,2]$, and
$\zeta(\lambda):\sigma_\varepsilon=[-2-\varepsilon,2+\varepsilon]\to\sigma_\varepsilon$ be some smooth function
such that $\inf_{\sigma_\varepsilon}\zeta'>0$. \\
Consider
\[H^{(\zeta)}(\bar\lambda)=-n\sum V(\zeta(\lambda_j))+\sum_{i\not=j}\log|\zeta(\lambda_i)-\zeta(\lambda_j)|+
\frac{2}{\beta}\sum\log\zeta'(\lambda_j).\]
It is evident that the correspondent partition function and all the marginal densities satisfy the relations
\begin{align*}Z_{n,\beta}^{(\zeta)}&:=\int e^{\beta H^{(\zeta)}/2}d\bar\lambda=Z_{n}[\beta,V],\\
p_{n,\beta}^{(m,\zeta)}(\lambda_1,\dots,\lambda_m)&:=(Z_{n,\beta}^{(\zeta)})^{-1}\int e^{\beta H^{(\zeta)}/2}
d\lambda_{m+1}\dots d\lambda_{n}\\&=p_{n,\beta}^{(m)}(\zeta(\lambda_1),\dots,\zeta(\lambda_m)).
\end{align*}
On the other hand,
\begin{align*}H^{(\zeta)}(\bar\lambda)=&-n\sum V(\zeta(\lambda_j))+\sum_{i\not=j}\log|\lambda_i-\lambda_j|\\
&+\sum_{i,j}\log\Big|\frac{\zeta(\lambda_i)-\zeta(\lambda_j)}{\lambda_i-\lambda_j}\Big|+
(\frac{2}{\beta}-1)\sum\log\zeta'(\lambda_j),\end{align*}
where we removed the condition $i\not=j$ in the third sum and add the correspondent terms to the forth sum.
Choose $\zeta(\lambda)$ from the equation
\begin{align}\label{eq_z}
\zeta'(\lambda)=\frac{\rho_{sc}(\lambda)}{\rho(\zeta(\lambda))},\quad\zeta(-2)=-2,\quad \mathrm{with}\quad\rho_{sc}(\lambda)=
\frac{\sqrt{4-\lambda^2}}{2\pi}.\end{align}
We will use this equation also in the form
\begin{align}\label{eq_z.1}
\rho(\zeta(\lambda))\zeta'(\lambda)=\rho_{sc}(\lambda).
\end{align}
\begin{lemma}\label{l:zeta}
$\zeta(\lambda)$ is a real analytic function in some $\sigma_{\tilde\varepsilon}$, and
$\zeta(2)=2$.
\end{lemma}
Without loss of generality we assume below that $\sigma_{\tilde\varepsilon}=\sigma_{\varepsilon}$.

For this choice of $\zeta$ write
\begin{align*}
\sum_{i,j} L^{(\zeta)}(\lambda_i,\lambda_j)=&\sum\eta_k\Big(\sum_j(\varphi_{k}(\lambda_j)-(\varphi_{k},\rho_{sc}))\Big)^2\\&+
2n\sum_{j}\sum_k\eta_k\varphi_{k}(\lambda_j)(\varphi_{k},\rho_{sc})
-n^2\sum_k\eta_k(\varphi_{k},\rho_{sc})^2\\=\Delta(\bar\lambda)+2n\sum_{j}\int &L^{(\zeta)}(\lambda_j,\mu)\rho_{sc}(\mu)d\mu-
n^2\int L^{(\zeta)}(\lambda,\mu)\rho_{sc}(\lambda)\rho_{sc}(\mu)d\lambda d\mu,
\end{align*}
where $(f,g):=\int_{\sigma_\varepsilon} fg d\lambda$. It is easy to see that for $\lambda\in\sigma$ in view of (\ref{eq_z.1})
\begin{align}\label{log_z}
2\int L^{(\zeta)}(\lambda,\mu)\rho_{sc}(\mu)d\mu=&
2\int\log|\zeta(\lambda)-\zeta(\mu)|\rho_{sc}(\mu)d\mu-2\int\log|\lambda-\mu|\rho_{sc}(\mu)d\mu\\
=&2\int\log|\zeta(\lambda)-\zeta(\mu)|\rho(\zeta(\mu))\zeta'(\mu)d\mu-\frac{\lambda^2}{2}
=V(\zeta(\lambda))-\frac{\lambda^2}{2}.
\notag\end{align}
On the other hand, the l.h.s. here is a real analytic function in $\sigma_{\varepsilon}$ and the r.h.s. is
also a real analytic function in $\sigma_{\varepsilon}$, hence (\ref{log_z}) is valid for
$\lambda\in\sigma_{\varepsilon}$. Similarly
\begin{align*}
\int L^{(\zeta)}(\lambda,\mu)\rho_{sc}(\lambda)\rho_{sc}(\mu)d\lambda d\mu=\mathcal{E}_{sc}-\mathcal{E}_V
=:-\Delta\mathcal{E}.
\end{align*}
Hence we finally obtain that our Hamiltonian  for $\bar\lambda\in\sigma_{\varepsilon}^n$ has the form (\ref{new_H})

\subsection{"Linearization" of the quadratic terms in (\ref{new_H})}
As it was mentioned in Section 1, in the case of real analytic $\zeta$, the eigenfunctions
$\{\varphi_k(\lambda)\}_{k=1}^\infty$ are analytic in the same domain as $\zeta$, and the eigenvalues $|\eta_k|\le e^{-kc}$.
Hence if we choose $M=[\log ^2n]$, then the remainder of the sum in (\ref{new_H}) will be less than
any negative degree of $n$ and will not have essential influence on the correlation functions.
Write for any $1\le k\le M$
\begin{align}\label{H-S}&\exp\Big\{\frac{\beta}{2}\eta_k\Big(\sum_j(\varphi(\lambda_j)-(\varphi_k,\rho_{sc}))\Big)^2\Big\}\\&=
\sqrt{\frac{\beta}{8\pi}}\int du_k\exp\Big\{\frac{\beta}{2}\Big(\sqrt{\eta_k}\Big(\sum_j(\varphi_k(\lambda_j)
-(\varphi_k,\rho_{sc}))\Big)u_k-u_k^2/4\Big)\Big\},
\notag\end{align}
where for $k\in I_+=\{k\le M:\eta_k>0\}$ we take an arithmetic square root, while for
$k\in I_-=\{k\le M:\eta_k<0\}$ $\sqrt{\eta_k}=i
\sqrt{|\eta_k|}$. We will write $\bar u=(u_1,\dots,u_M)$.
%, where $\bar u^{+}=\{u_k\}_{k\in I_+}$ and
%$\bar u^{-}=\{u_k\}_{k\in I_-}$.
Substituting this integrals in (\ref{new_H}) and integrating first with respect
to $\bar\lambda$, we get (\ref{int_Phi}) with
\begin{align}\label{h_u}h_{\bar u}(\lambda)=\sum_{k=1}^M\sqrt{\eta_k}\varphi_k(\lambda)u_k
+(\frac{2}{\beta}-1)\log\zeta'(\lambda),\quad \dot h_{\bar u}(\lambda)=h_{\bar u}-(h_{\bar u},\rho_{sc}).\end{align}

\subsection{Integration with respect to $\bar u$}
The first our step is
 to get rid in (\ref{int_Phi}) from the domain of $\bar u$, where $\bar u$ is big, proving that the correspondent contribution
in the integral (\ref{int_Phi}) is small.
Set
\begin{align}\notag
&K^+_{jk}=\eta_j^{1/2}\eta_k^{1/2}(\bar D_\sigma\varphi_k,\varphi_j),\;j,k\in I_+,\quad\\
&K^-_{jk}=|\eta_j|^{1/2}|\eta_k|^{1/2}(\bar D_\sigma\varphi_k,\varphi_j),\;j,k\in I_-,\notag\\
&U_1=\{\bar u: (K^+\bar u,\bar u)+(K^-\bar u,\bar u)\le k_*\log \varepsilon_n^{-1}\},
\label{U_1}\end{align}
where $\varepsilon_n$ is given by (\ref{VV2}) and $k_*$ is some absolute constant which  will be chosen later.
\begin{lemma}\label{l:conv}
There exists $n$-independent $\delta>0$ such that
\begin{align}\label{b_L}
K^+<1-\delta.
\end{align}
Moreover, if $U_1^c$ is a complement of $U_1$ of (\ref{U_1}), then
\begin{align}
&\Big(\frac{\beta}{8\pi}\Big)^{M/2}
\int_{U_1^c}d\bar u e^{-\beta(\bar u,\bar u)/8}|\langle \Phi_ke^{\beta\mathcal{N}_n[\dot h_u]/2}\rangle_{*,n}|
  \le \varepsilon_n^{\tau k_*},
\label{conv.2}\end{align}
where $\varepsilon_n$ is given by (\ref{VV2}), $k_*$ - by (\ref{U_1}), and $\tau>0$ is some fixed number, depending on $\delta$ in (\ref{b_L}).
\end{lemma}
The proof of Lemma \ref{l:conv}  is partially based on the following assertion
\begin{lemma}\label{l:re_h}
Let  $h(\lambda)$ be a real analytic function such that $||h'(\lambda)||_2,\,||h^{(16)}(\lambda)||_2\le \log^sn$
with some $n$-independent positive $s$. Then
\begin{align}\label{b_re_h}
&\Big|\frac{\langle \Phi_ke^{\beta\mathcal{N}_n[\dot h]/2}\rangle_{*,n}}{\langle e^{\beta\mathcal{N}_n[\dot h]/2}\rangle_{*,n}}-
\langle\Phi_k(\lambda)\rangle_{*,n}\Big|\le
C (n^{-\kappa}+\varepsilon_n),
\end{align}
where  $\varepsilon_n$ is the same as in (\ref{VV2}), and $\kappa>0$.
\end{lemma}
In particular,
Lemma \ref{l:re_h} and (\ref{VV1}) for real $h$ imply the bound which we need in the proof of Lemma \ref{l:conv}:
\begin{align}\label{b_Phi}
\Big|{\langle\Phi_ke^{\beta\mathcal{N}_n[\dot h]/2}\rangle_{*,n}}\Big|
\le C_{\Phi_k}{\langle e^{\beta\mathcal{N}_n[\dot h]/2}\rangle_{*,n}}.
\end{align}
Now let us prove (\ref{**}). As it was mentioned above, for  real  $h_u$ (\ref{**}) follows from
 Lemma \ref{l:re_h} (see (\ref{b_re_h})).
To extend (\ref{b_re_h}) to the
complex valued $h_u$, we use the last lemma:
\begin{lemma}\label{l:3}
Let the analytic in $t\in D=\{t:|t|\le \log^{1/2}\varepsilon_n^{-1},\Im t\ge 0\}$ functions $F_n$ satisfy two bounds:
\begin{align}\label{l3.1}
 &|F_n(t)|\le C_1\varepsilon_n e^{t^2/2},\quad
-\log^{1/2}\varepsilon_n^{-1}\le t\le \log^{1/2} \varepsilon_n^{-1},\quad \varepsilon_n<1   ,\\
&|F_n(t)|\le C_2e^{(\Re t)^2/2}, \quad t\in D.
\notag\end{align}
Then the inequality
\begin{equation}\label{l3.2}
   | F_n(t)|\le C\varepsilon_n^{1/2}|e^{t^2/2}|
\end{equation}
holds  for $t\in D':=\frac{1}{6}D$ with $C= C_1^{3/4}C_2^{1/4}$.
\end{lemma}

Denote
\begin{align}\label{X,d}
&X_1=\mathcal{N}_n[\Re h_{\bar u}],\quad X_2=\mathcal{N}_n[\Im h_{\bar u}],\\
&d_{11}=\frac{\beta}{4}(D\Re h_{\bar u},\Re h_{\bar u}),\quad
d_{12}=\frac{\beta}{4}(D\Re h_{\bar u},\Im h_{\bar u}),
\quad d_{22}=\frac{\beta}{4}(D\Im h_{\bar u},\Im h_{\bar u}),
\notag\end{align}
and use Lemma \ref{l:3} for
\[
F_n(t)=\frac{\langle \Phi_k(\bar\lambda)e^{\beta(X_1-X_2d_{12}/d_{22}+tX_2/d_{22}^{1/2})/2}\rangle_{*,n}}
{\langle e^{\beta(X_1-X_2d_{12}/d_{22})/2}\rangle_{*,n}}-e^{t^2/2}\langle\Phi_k(\bar\lambda)\rangle_{*,n},
\]
Lemma \ref{l:re_h}, (\ref{Joh}), and (\ref{VV1}) guarantee that $F_n(t)$ satisfy (\ref{l3.1}).
Take $t^*=id_{22}^{1/2}+d_{12}/d_{22}^{1/2}$. For  $k_*\le\frac{1}{6}$ in (\ref{U_1}), $t\in \frac{1}{6}D$, since
\begin{align} \label{*.2}
&k_*\log \varepsilon_n^{-1}\ge d_{11}+d_{22}\ge d_{22}+d_{12}^2/d_{22}=|id_{22}^{1/2}+d_{12}/d_{22}^{1/2}|=|t^*|.
\end{align}
In addition, by (\ref{Joh}) and (\ref{X,d})
\[
e^{t^2/2}\langle e^{\beta(X_1-X_2d_{12}/d_{22})/2}\rangle_{*,n}=e^{d_{11}+2id_{12}-d_{22}}(1+O(n^{-\kappa})),
\]
 hence  Lemma \ref{l:3} yields (cf (\ref{**}))
\begin{align}\label{t1.l}
\Big|\langle \Phi_k(\bar\lambda)e^{\beta(X_1+iX_2)/2}\rangle_{*,n}-
e^{d_{11}+2id_{12}-d_{22}}\langle\Phi_k(\bar\lambda)\rangle_{*,n}
\Big|
\le C\varepsilon_n^{1/2}\Big|e^{d_{11}+2id_{12}-d_{22}}\Big|.
\end{align}
Applying this inequality first for $\Phi\equiv 1$, we get
that
\[\langle e^{\beta(X_1+iX_2)/2}\rangle_{*,n}=e^{d_{11}+2id_{12}-d_{22}}(1+O(\varepsilon_n^{-1/2})),\]
and then, substituting the last relation in (\ref{t1.l}), we obtain (\ref{**}).
Integrating (\ref{**}) in $U_1$ we complete the proof of Theorem \ref{t:1}.

$\square$

\section{Proofs of the auxiliary results}
\textit{Proof of Lemma \ref{l:zeta}.}
The fact that $\zeta(2)=2$ follows from the relation (\ref{eq_z.1}) and the fact that
\[\int_{-2}^2\rho(\lambda) d\lambda=\int_{-2}^2\rho_{sc}(\lambda) d\lambda=1.\]
The analyticity in all internal points of $(-2,2)$ follows from the analyticity of $P$ (see (\ref{P})).
Hence we are left to prove that $\zeta(\lambda)$ is analytic in some neighborhood of $\lambda=\pm2$.

Consider, e.g., $\lambda=-2$. To simplify formulas,  we make the change $x=\lambda+2$. Let us sick solution
in the form
\[\zeta(x)+2=P_0^{-2/3}x(1+\zeta_0(x)),\quad \zeta_0(0)=0.\]
where $P_0:=P(-2)\not=0$. Then
\[
P(\zeta)=P_0(1+\zeta \tilde P(\zeta))=P_0(1+x(1+\zeta_0)P_1(\zeta_0,x))\]
 ($P_1(\zeta_0,x)$ is analytic in both variables)
and (\ref{eq_z}) can be written as
\begin{align*}
(x\zeta_0)'=&\frac{\sqrt{4-x}}
{(1+x(1+\zeta_0) P_1(\zeta_0,x))\sqrt{(1+\zeta_0)(4-P_0^{-2/3}x(1+\zeta_0(x)))}}-1\\
:=&F(x,\zeta_0)=(xF_0(x)+\zeta_0F_1(x)+\zeta_0^2F_2(x,\zeta_0)),
\end{align*}
where  we used the fact that the r.h.s. of the first line is analytic in $x,\zeta_0$ at the point $(0,0)$, hence $F_0,F_1,F_2$
are analytic at $(0,0)$, moreover   the r.h.s.
is 0 at this point. Note that
\[F_0(x)=\frac{\partial F}{\partial \zeta_0}\Big|_{\zeta_0=0}=
\frac{\partial\zeta}{\partial \zeta_0}\frac{\partial F}{\partial \zeta}\Big|_{\zeta_0=0}=
xP_0^{-2/3}\frac{\partial }{\partial \zeta}\frac{\sqrt{x(4-x)}}{P(\zeta)\sqrt{4-\zeta^2}}\Big|_{\zeta_0=0}\]
 can not be identically  zero,
if  $P(x)\not\equiv 1$. Moreover, $F_1(x)=-\frac{1}{2}+xF_{11}(x)$. Thus, the equation can be written in the form
\begin{align*}
\zeta_0'=&-\frac{3}{2}\frac{\zeta_0}{x}+F_0(x)+\zeta_0F_{11}(x)+\frac{\zeta_0^2}{x} F_2(x,\zeta_0),\quad
 F_0(x)=x^mF_{0,m}(x),\quad F_{0,m}(0)\not=0,
\end{align*}
where $m$ could be 0 or any positive integer. It is evident that if we sick $\zeta_0=\sum_{k=1}\zeta_kx^k$,
then the equation above gives us the recursive system
\begin{align}\label{rec}
(k+\frac{3}{2})\zeta_k=P_k(\zeta_1,\dots,\zeta_{k-1}),
\end{align}
where $P_k$ is a polynomial of $\zeta_1,\dots,\zeta_{k-1}$ with coefficients depending on the Taylor coefficients of $F_0,F_{11},F_2$. This system
always has a solution, the  only problem is to check that the corresponding series is convergent, i.e. to find the upper
bounds for $|\zeta_k|$. It is clear that if we replace all coefficients of $P_k$ by something bigger, then the solution
$\zeta_k$ becomes bigger and similarly one can replace $(k+\frac{3}{2})\to 2$. If
  $F_0,F_{11},F_2$ are  analytic functions in $x,\zeta_0$ for $|x|,|\zeta_0|\le\varepsilon_1$, their Taylor coefficients
are less than the corresponding coefficients of the functions $Ax^m(\varepsilon_1-2x)^{-1}$, $A(\varepsilon_1-2x)^{-1}$
and $A(\varepsilon_1-2x)^{-1}(\varepsilon_1-2\zeta_0)^{-1}$, where $A$ is a sufficiently big number.
Hence, the coefficients  solving (\ref{rec}) are less
than the coefficients of the solution of the algebraic equation
\begin{align*}
 \frac{2\zeta_0}{x}=\frac{A(x^m+\zeta_0)}{(\varepsilon_1-2x)}+\frac{A\zeta_0^2}{x(\varepsilon_1-2\zeta_0)(\varepsilon_1-2x)}
\end{align*}
One can easily check that the solution of this quadratic equation is an analytic function
at $x=0$, hence the coefficients  solving (\ref{rec})
give us an analytic function at $x=0$.

$\square$

\textit{Proof of Lemma \ref{l:conv}}. We start from the technical proposition, whose proof is given after the proof
 of Lemma \ref{l:conv}.
\begin{proposition}\label{p:L_+} Set $L^{(\zeta)}_{l}(x,y):=\frac{\partial^{l}}{\partial x^{l}}L^{(\zeta)}(x,y)$,
and denote $L^{(\zeta)}_{l}$ the integral operator in $L_2[\sigma_\varepsilon]$ with this kernel.
Then
\begin{align}\label{pL.1}
\sum |\eta_k|(D_\sigma\varphi_k,\varphi_k)&\le C(\varepsilon)
\Big(\mathrm{Tr\,}L^{(\zeta)}_{2}L^{(\zeta)*}_{2}+\mathrm{Tr\,}L^{(\zeta)}_{1}L^{(\zeta)*}_{1}\Big)^{1/2},\\
\sum |\eta_k|(\varphi_k^{(l)},\varphi_k^{(l)})_{\varepsilon/2}&\le
C_l(\varepsilon)\Big(\mathrm{Tr\,}L^{(\zeta)}_{2l+1}L^{(\zeta)*}_{2l+1}
+\mathrm{Tr\,}L^{(\zeta)}_{1}L^{(\zeta)*}_{1}\Big)^{1/2}.
\label{pL.2}\end{align}
Here and below we denote by $(.,.)_{\varepsilon/2}$ the standard scalar product in $L_2[\sigma_{\varepsilon/2}]$.
\end{proposition}
%Let $L^{(\zeta)}_+,L^{(\zeta)}_- $ be  positive and negative parts of the operator $L^{(\zeta)}$.
The Schwartz inequality and (\ref{pL.1}) imply that $K_+$ is a Hilbert-Schmidt matrix,
hence, its eigenvalues $\mu_k\to 0$, and therefore $\mu_k\le\frac{1}{2}$
 for all  $k$ except may be  a finite  set $\{k_1,\dots,k_\ell\}:=I$. Moreover, the definition of $K^+$, the standard
properties of the operator norm and (\ref{DL}) yield
 for $\varphi=\sum u_k\varphi_k$
\begin{align*}
&(K^+u,u)=((L^{(\zeta)}_+)^{1/2}\overline D_\sigma(L^{(\zeta)}_+)^{1/2}\varphi,\varphi)
\le ||(L^{(\zeta)}_+)^{1/2}\overline D_\sigma(L^{(\zeta)}_+)^{1/2}||\\=&
||\overline D_\sigma^{1/2}L^{(\zeta)}_+\overline D_\sigma^{1/2}||\le
||\overline D_\sigma^{1/2}L_\sigma\overline D_\sigma^{1/2}||\le 1,
\end{align*}
where $L^{(\zeta)}_+$ is a  positive  part of the operator $L^{(\zeta)}$ and $L_\sigma$ is defined in (\ref{DL}).  Note that since the definition of $D_\sigma$ includes projection $\Pi$
on the interval $\sigma$, we can use the inequality $\Pi L^{(\zeta)}_+\Pi \le -L_\sigma$.
Hence, $((L^{(\zeta)}_+)^{1/2}\overline D_\sigma(L^{(\zeta)}_+)^{1/2} \varphi,\varphi)=1$ for some $\varphi$ only if
for $\tilde\varphi=(L^{(\zeta)}_+)^{1/2}\varphi$ we have
\begin{align}\label{b_DL}
L^{(\zeta)}_+\overline D_\sigma \tilde\varphi=\tilde\varphi\Rightarrow
(\overline D_\sigma L^{(\zeta)}_+\overline D_\sigma \tilde\varphi,\tilde\varphi)
=(\overline D_\sigma\tilde\varphi,\tilde\varphi)=-(\overline D_\sigma L_\sigma\overline D_\sigma \tilde\varphi,\tilde\varphi),
\end{align}
where the last equality follows from (\ref{DL}).
On the other hand, $\Pi L^{(\zeta)}_+\Pi \le -L_\sigma$, hence the above equality is possible only if
$L^{(\zeta)}_+\overline D_\sigma \tilde\varphi=L\overline D_\sigma \tilde\varphi$. But
$\Pi L^{(\zeta)}_+\Pi \phi=-L_\sigma\phi$, only if
\[\int_\sigma\log|\zeta(\lambda)-\zeta(\mu)|^{-1}\phi(\mu)\phi(\lambda)d\lambda d\mu=0,\]
which contradicts to the positivity  of the operator with the kernel $\log|\zeta(\lambda)-\zeta(\mu)|^{-1}$. Thus,
$\sup_{k\in I}\mu_k\le 1-\delta_1$ with $\delta_1>0$.
Choosing $\delta=\min\{\delta_1,\frac{1}{2}\}$ we obtain (\ref{b_L}).

To prove (\ref{conv.2}), we   denote
\begin{align}\label{sc_prod}
(A_0\bar u,\bar u):=\sum_{k,j\le M}u_ku_j|\eta_k|^{1/2}|\eta_j|^{1/2} ( \varphi_k^{(16)}, \varphi_j^{(16)})_{\varepsilon/2}
\end{align}
(we need the 16th derivative here to control the 16th derivative of $h_u$ in Lemma \ref{l:re_h}) and set
\begin{align}\label{U}
U_2=&\{\bar u:(A_0\bar u,\bar u)\le
\log^2 n\,\wedge\,
(K^+\bar u,\bar u) +(K^-\bar u,\bar u)\ge {k_*}\log\varepsilon_n^{-1}\},\\
U_3=&\{\bar u:\log^2n\le
(A_0\bar u,\bar u)\le n\log^2n\},\notag\\
U_4=&\{\bar u: n\log^2n\le (A_0\bar u,\bar u)\le C_*n^2\},\notag\\
U_5=&\{\bar u: C_*n^2\le (A_0\bar u,\bar u)\}
\notag\end{align}
with sufficiently large $n$-independent $C_*$. One can see easily that
\[
U_1^c\subset U_2\cup U_3\cup U_4\cup U_5,
\]
Below we will often use the following evident statement
\begin{proposition}\label{r:int} For any semi-infinite  matrix  $A>0$ such that $\sum_{i=1} A_{ii}<\infty$
and   $||A||<1-\delta$ ($\delta>0$)
\begin{align}\label{int_*}
\Big(\frac{\beta}{8\pi}\Big)^{M/2}\int e^{-\beta(\bar u,\bar u)/8}e^{\beta\sum (A\bar u,\bar u)/8} d\bar u\le C
\end{align}
with some $M$-independent $C$.\end{proposition}
In particular, (\ref{int_*}) is true for $(A\bar u,\bar u)=(K^+\bar u,\bar u)+\tau(A_0\bar u,\bar u)$,
with sufficiently small $\tau>0$, since we proved above that $||K^+||\le 1-\delta$ and
 Proposition \ref{p:L_+}  guarantees that $\sum_{i=1} A_{ii}<\infty$.

By the Schwartz  inequality,
\begin{align*}
|&\Re h_u(\bar\lambda)|=\Big|\sum_{k\in I_+}u_k\sqrt\eta_k\varphi_k(\lambda)\Big|
\le||\bar u||
\Big|\sum_{k\in I_+}\eta_k|\varphi_k(\lambda)|^2\Big|\\ &\le||\bar u||(L^{(\zeta)}_+(\lambda,\lambda))^{1/2}\le
C||\bar u||.\end{align*}
Hence
\begin{align*}
\langle \Phi_k(\bar\zeta)e^{\beta\mathcal{N}_n[\dot h_u]/2}\rangle_{*,n}\le
(nC)^ke^{\beta \max_{\lambda}\{n|\Re h_u(\bar\lambda)|\}/2}\le
 e^{nC_1(1+||\bar u||)},
\end{align*}
where we used the trivial bound
\begin{align}\label{tr_b}|\Phi_k|\le (nC)^k.\end{align}
Then, using the fact that the matrix $A_0$ defined by the quadratic form $(A_0\bar u,\bar u)$ is bounded
(in view of  Propositions \ref{p:L_+}), we have  in $U_5$
$||\bar u||^2||A_0||\ge(A_0\bar u,\bar u)>n^2C_*$. Hence  for sufficiently large $C_*$ the integral
\begin{align*}
&\int_{U_5} du e^{-\beta(\bar u,\bar u)/8}\langle \Phi_k(\bar\zeta)e^{\beta\mathcal{N}_n[\dot h_u]/2}\rangle_{*,n}\\\le
&\int_{||\bar u||^2>n^2C_*/||A_0||} d\bar u e^{-\beta(\bar u,\bar u)/8}
e^{nC(1+||\bar u||)}\le e^{-Cn^2}.
\end{align*}
For $u\in U_4$
\begin{align}\label{b_s'}
n^{-2}\int|\Re h_u^{(16)}(\lambda)|^2d\lambda=n^{-2}(A_0\bar u,\bar u)\le C_*.
\notag\end{align}
Thus  $n^{-1} \Re h_u(\lambda)$ is a H\"{o}lder function for $u\in U_4$, and we can use the result
of \cite{BPS:95}, according to which
\begin{align*}&Z_{ n}[V^{*}-n^{-1}\Re h_u]
\\&\le
\exp\Big\{
\frac{\beta n^2}{2}\max_{m\in\mathcal{M}_1^+[\sigma_{\varepsilon/2}]}\{L[m,m]-(m,V^{*}
-n^{-1} h_u)\}+Cn\log n\Big\},
\end{align*}
where $\mathcal{M}_1^+[\sigma_{\varepsilon/2}]$ is a set of positive unit measures with supports belonging to
$\sigma_{\varepsilon/2}$.
Since
\[-V^{*}(\lambda)\le -2L[\rho_{sc}](\lambda),\quad \lambda
\in\sigma_{\varepsilon},\]
we have
\begin{align}\notag
&\max_{m\in\mathcal{M}_1^+[\sigma_{\varepsilon}]}\{L[m,m]-(m,V^{*}
-n^{-1}\Re h_u)\}\\\le&\max_{m\in\mathcal{M}_1^+[\sigma_{\varepsilon}]}
\{L[m,m]-(m,2\mu_\alpha^{-1}L[\rho_{sc}]
-n^{-1}\Re h_u)\}\notag\\
\le&\max_{m\in\mathcal{M}_1[\sigma_{\varepsilon}]}\{L[m,m]-(m,2L[\rho_{sc}]
-n^{-1}\Re h_u)\}=:E(\bar u).\label{b_E}\end{align}
Here $\mathcal{M}_1[\sigma_{\varepsilon/2}]$ is a set of all signed unit measures with supports belonging to
$\sigma_{\varepsilon/2}$. It is easy to see that, if we remove the condition of positivity of measures,
then the maximum point $\rho_{1}$ is uniquely defined by the conditions:
\[2L[\rho_{1}](\lambda)-2L[\rho_{sc}](\lambda)
-n^{-1}\Re h_u(\lambda)=\mathrm{const},\quad\lambda\in
 \sigma_{\varepsilon},\quad\int_{\sigma_{\varepsilon}}\rho_{1}=1.\]
Hence $\rho_{1}=\rho_{sc}+
\frac{1}{2}D_{\sigma_{\varepsilon}}\Re h_u$ and the r.h.s. of (\ref{b_E}) takes the form
\begin{align*}
E(\bar u)=-L[\rho_{sc},\rho_{sc}]+
\frac{n^{-2}}{4}(\bar D_{\sigma_{\varepsilon}}\Re h_u,\Re h_u)+n^{-1}
(\dot\Re  h_u,\rho_{sc}).
\end{align*}
 But by the definition of $\dot h_u$ $(\dot  h_u,\rho_{sc})=0$.
Hence
\[E(\bar u)=-L[\rho_{sc},\rho_{sc}]+
\frac{n^{-2}}{4}(K^+_\varepsilon\bar u,\bar u)
+O(n^{-1}\log n),\]
where $K^+_\varepsilon$ is defined by the same way as $K^+$ (see (\ref{U_1})), but with $\bar D_\sigma$ replaced by
$\bar D_{\sigma_{\varepsilon/2}}$.
These relations and (\ref{tr_b})  yield
\begin{align*}\langle \Phi_k(\bar\zeta)e^{\beta\mathcal{N}_n[\dot h_u]/2}\rangle_{*,n}\le& (Cn)^k
e^{\frac{\beta}{8}(K^+_\varepsilon\bar u,\bar u)+O(n\log n)}
%&\le&\exp\Big\{\frac{\beta}{8}(1-\delta_1)(u_+, u_+)+O(n\log n)\Big\}.
\end{align*}
Then the Chebyshev inequality for sufficiently small  $\tau$ and   (\ref{int_*}) yield
\begin{align}\label{U_4}
&\Big(\frac{\beta}{8\pi}\Big)^{M/2}\int_{U_4}e^{-\beta(\bar u,\bar u)/8}
\langle \Phi_k(\bar\zeta)e^{\beta\mathcal{N}_n[\dot h_u]/2}\rangle_{*,n} d\bar u\\
\le&\Big(\frac{\beta}{8\pi}\Big)^{M/2}e^{O(n\log n})
\int e^{-\beta((I-K^+_\varepsilon)\bar u,\bar u)/8+\tau((A_0\bar u,\bar u)-n\log^2n)}d\bar u\le e^{- \tau n\log^2n/2}.
\notag\end{align}
For $u\in U_3$ (\ref{tr_b}) and (\ref{Joh})  imply
\begin{align}\label{U_3}
\langle \Phi_k(\bar\zeta)e^{\beta\mathcal{N}_n[\dot h_u]/2}\rangle_{*,n}&\le (Cn)^k\exp\Big\{
\frac{\beta}{4}(K^+\bar u,\bar u)+(\bar u,\bar q)+C\\
&+Cn^{-1}||\Re h_u^{(6)}||_2^{3}+||\Re h_u^{'}||_2^{3}\big)\Big\}\notag\\
&\le \exp\Big\{\frac{\beta}{4}(K^+\bar u,\bar u)+
n^{-1/3}(A_0\bar u,\bar u)+C\Big\},
\notag\end{align}
where we used that the vector $\bar q=(q_1,\dots,q_M)$, $q_k=\eta_k^{1/2}\beta( \bar D_\sigma\varphi_k,\zeta')+\eta_k^{1/2}(\nu_\beta,\varphi_k)$
is bounded (it is easy to check by the Schwartz inequality) and  that
\begin{align*}
&||\Re h_u^{(l)}||_2^2\le C_l(1+||\Re h_u^{(16)}||_2^2)=C(1+(A_0\bar u,\bar u)),\quad l=1,6,\\
& n^{-1}(A_0\bar u,\bar u)^{3/2}\le
\log n n^{-1/2}(A_0\bar u,\bar u)\le n^{-1/3}(A_0\bar u,\bar u),\quad\bar u\in U_3.
\end{align*}
 Then, similarly to (\ref{U_4}), the Chebyshev inequality with sufficiently small $\tau$ yields
\begin{align*}
&\Big(\frac{\beta}{8\pi}\Big)^{M/2}\int_{U_3}e^{-\beta(\bar u,\bar u)/8}
\langle \Phi_k(\bar\zeta)e^{\beta\mathcal{N}_n[\dot h_u]/2}\rangle_{*,n}d\bar u\\
&\le \Big(\frac{\beta}{8\pi}\Big)^{M/2}(Cn)^k
\int e^{-\beta((I-K^+)\bar u,\bar u)/8+(\bar u,\bar q)+(\tau+n^{-1/3})(A_0\bar u,\bar u)-\tau\log^2n}du\le e^{- \tau\log^2n/2}.
\end{align*}
Finally,   using the bound (\ref{b_Phi}) for  $\bar u\in U_2$
and again the Chebyshev inequality with sufficiently small $\tau$,
we obtain the bound, finishing the proof of the lemma:
\begin{align}\label{U_2}
&\Big(\frac{\beta}{8\pi}\Big)^{M/2}\int_{U_2}e^{-\beta(\bar u,\bar u)/8}
\Big|\langle \Phi_k(\bar\zeta)e^{\beta\mathcal{N}_n[\dot h_u]/2}\rangle_{*,n}\Big| d\bar u\\
\le&C_{\Phi_k}\Big(\frac{\beta}{8\pi}\Big)^{M/2}\int \int d\bar u e^{-\beta((I-K^+)\bar u,\bar u)/8+(\bar u,\bar q)}\notag\\
&\quad\cdot e^{n^{-1/3}(A_0\bar u,\bar u)+\tau((K^+\bar u,\bar u)+(K^-\bar u,\bar u)-k_*\log \varepsilon_n^{-1})}
\le C_3e^{\tau k_*\log \varepsilon_n}.
\notag\end{align}
% Lemma \ref{l:conv} is proved.
$\square$

\textit{Proof of Proposition \ref{p:L_+}}. By  (\ref{DL})
\begin{align}\notag
\sum&|\eta_k|(\bar D_\sigma\varphi_k,\varphi_k)=\sum|\eta_k|(\bar D_\sigma^2\varphi_k,(-L_\sigma)\varphi_k)\\&\le
\Big(\sum|\eta_k|^2(\bar D_\sigma^2\varphi_k,\bar D_\sigma^2\varphi_k)\Big)^{1/2}
\Big(\sum( L_\sigma^2\varphi_k,\varphi_k)\Big)^{1/2}
\label{pL.3}\\
&\le C\Big(\int dxdy
\Big(\sum\eta_k\bar (D_\sigma^2\varphi_k)(x)\varphi_k(y)\Big)^2=C\int |\bar D_{\sigma}^2L^{(\zeta)}(x,y)|^2dxdy.
\notag\end{align}
The last factor in the second line here is bounded since $(-L_\sigma)$ is a Hilbert-Schmidt operator.
In addition, according to (\ref{bar_D.1}), for any $h$ such that $h^{(m)}\in L_2[\sigma]$
\begin{align}\label{b_D}
(\bar D_\sigma^{2m}h,h)=&C\int_0^\pi \Big|\frac{d^{m}}{d\theta^m}h(2\cos\theta)\Big|^2d\theta=
C\int_{-2}^2 \Big|\Big(|X^{1/2}(x)|\frac{d}{dx}\Big)^mh(x)\Big|^2\frac{dx}{|X^{1/2}(x)|}\\
\le&C\sum_{p=1}^m ||h^{(p)}||_2^2\le \tilde C(||h^{(m)}||_2+||h^{'}||_2).
\notag\end{align}
Here we  used that for $p\le m-1$, $||h^{(p)}||_\infty\le C(||h^{(p)}||_2+||h^{(p+1)}||_2)$,
hence the last integral in (\ref{b_D}) is convergent. Applying (\ref{b_D}) to the r.h.s. of (\ref{pL.3})
we obtain (\ref{pL.1}).

To obtain (\ref{pL.2}), we write
\begin{align*}
\int_{\sigma_{\varepsilon/2}}|h^{(l)}(x)|^2dx=\int_{|\cos\theta|\le \frac{1+\varepsilon/2}{1+\varepsilon}}
\Big|\Big(\frac{1}{\sin\theta}\frac{d}{d\theta}\Big)^lh(2(1+\varepsilon)\cos\theta)\Big|^2\sin\theta d\theta\\
\le \tilde C_l(\varepsilon)\int_0^\pi\Big|\Big(\frac{d}{d\theta}\Big)^lh(2(1+\varepsilon)\cos\theta)\Big|^2d\theta=
C_l'(\varepsilon)(\bar D_{\sigma_\varepsilon}^{2l}h,h).
\end{align*}
Hence, similarly to (\ref{pL.3})-(\ref{b_D}),
\begin{align*}
\sum|\eta_k|(\varphi_k^{(l)},\varphi_k^{(l)})_{\varepsilon/2}\le
C_l'(\varepsilon)\sum|\eta_k|(\bar D^{2l}_{\sigma_\varepsilon}\varphi_k,\varphi_k)=
\sum|\eta_k|(\bar D^{2l+1}_{\sigma_\varepsilon}\varphi_k,(-L_{\sigma_\varepsilon})\varphi_k)\\
\le
 CC_l'(\varepsilon)\Big(\sum|\eta_k|^2
(\bar D_{\sigma_\varepsilon}^{2l+1}\varphi_k,\bar D_{\sigma_\varepsilon}^{2l+1}\varphi_k)\Big)^{1/2}
\le C_l(\varepsilon)(\mathrm{Tr\,}L^{(\zeta)}_{2l+1}L^{(\zeta)*}_{2l+1}+\mathrm{Tr\,}L^{(\zeta)}_{1}L^{(\zeta)*}_{1})^{1/2}.
\end{align*}
$\square$

\textit{Proof of Lemma \ref{l:re_h}}.
The idea is to consider
\begin{align}\label{V_h}
V_h= \frac{\lambda^2}{2}+\dfrac{1}{n} h(\lambda)-c_1\frac{\lambda}{n}-c_2\frac{\lambda^2}{2n}:=
V_*+\frac{1}{n}\tilde h
\end{align}
with some appropriate $c_1$ and $c_2$ as a new potential and to apply
to it the above procedure with the change of variables. But since it is possible  only for the potentials
whose support of the equilibrium measure is $[-2,2]$, we need to have two equalities:
\[
\int_{\sigma}\frac{V_h'(\lambda)d\lambda}{X^{1/2}_\sigma(\lambda)}=0,\quad \pi^{-1}
\int_{\sigma}\frac{V_h'(\lambda)\lambda d\lambda}{X^{1/2}_\sigma(\lambda)}=1.
\]
Here the first equality is a necessary condition to have a bounded solution of the singular integral
equation which can be obtained by the differentiation of (\ref{cond_rho}), and the second equality provides the condition that
the integral of the corresponding density $\rho_h$ is 1. Thus we have to choose
\begin{align*}
c_1(h)=\pi^{-1}\displaystyle\int\dfrac{h'(\lambda)d\lambda}{X^{1/2}(\lambda)},
\quad c_2(h)=\pi^{-1}\displaystyle\int\dfrac{\lambda h'(\lambda)d\lambda}{2X^{1/2}(\lambda)}.
\end{align*}
Solving equation (\ref{eq_z}) for $V_h$, we obtain uniformly in $\lambda\in\sigma_\varepsilon$
\begin{align}\label{z_h}
\zeta_h=\lambda-\frac{1}{n}\tilde\zeta_h,\quad
\tilde\zeta'_h(\lambda)=\frac{\rho_{\tilde h}(\lambda)}{\rho_{sc}(\lambda)}+O(n^{-1}||h''||_2^2).
\end{align}
where $\rho_{\tilde h}$ is the equilibrium density, corresponding to $\tilde h$. According to (\ref{P}),
\[
\frac{\rho_{\tilde h}(\lambda)}{\rho_{sc}(\lambda)}=\int_{\sigma}\frac{\tilde h'(\mu)d\mu}
{(\lambda-\mu)X_\sigma^{1/2}(\mu)},
\]
hence, by the assumptions of the lemma $||\zeta_h^{(14)}||\le C\log^s n$.
Then the correspondent compact operator kernel $L^{(\zeta_h)}(\lambda,\mu)$ has the form
\[
L^{(\zeta_h)}(\lambda,\mu)=\log \frac{\zeta_h(\lambda)-\zeta_h(\mu)}{\lambda-\mu}=
\log\Big(1+\frac{1}{n}\frac{\tilde\zeta_h(\lambda)-\tilde\zeta_h(\mu)}{\lambda-\mu}\Big)
=\frac{1}{n}K_h(\lambda,\mu),
\]
where $K_h(\lambda,\mu)=K_h(\mu,\lambda)$ and for any $\mu$ $K_h(.,\mu)$ is a real analytic function
bounded by $C\log^s n$.
Let $\{\kappa_k,\varphi_{h,k}\}$ be eigenvalues and eigenvectors of $K_h$.
Then we obtain that the $k$th correlation function of the Hamiltonian with the potential (\ref{V_h})
 at the point $(\lambda_0+x_1/n\rho(\lambda_0),\dots,
\lambda_0+x_k/n\rho(\lambda_0))$ coincides with that at the point
$\big(\zeta_h(\lambda_0+x_1/n\rho(\lambda_0)),\dots,\zeta_h(\lambda_0+x_k/n\rho(\lambda_0))\big)$
for the Hamiltonian (cf (\ref{new_H}))
\begin{align}\notag
H^{(\zeta_h)}(\bar\lambda)=&-n\sum \Big(\frac{\lambda_i^2}{2}\Big(1-\frac{c_2}{n}\Big)
-\frac{c_1}{n}\lambda_i\Big)
+\sum_{i\not=j}\log|\lambda_i-\lambda_j|\\
&+\frac{1}{n}\sum_{k=1}^\infty\kappa_k\Big(\sum_j\Big((\varphi_{h,k}(\lambda_j)-(\varphi_{h,k},\rho_{sc})\Big)\Big)^2
+\frac{1}{n}(\frac{2}{\beta}-1)\sum\zeta^{(1)}(\lambda_i),
\label{H_h}\end{align}
where $\zeta_h^{(1)}(\lambda_i)=n\log(1+n^{-1}\tilde\zeta'_h)$.
Taking $M=[\log^2n]$, we obtain like before that we can restrict the summation
 above by $k=M$. Hence we get similarly to (\ref{H-S})
\begin{align*}
\frac{\langle\Phi_ke^{\beta\mathcal{N}_n[h]/2}\rangle_{*,n}}
{\langle e^{\beta\mathcal{N}_n[h]/2}\rangle_{*,n}}=&I_n^{-1}[\beta,\zeta_h]\Big(\frac{\beta}{8\pi}\Big)^{M/2}
\int e^{-\beta(\bar u,\bar u)/8}d\bar u
\langle\tilde\Phi_ke^{\beta\mathcal{N}_n[c_1\ell_1+c_2\ell_2+n^{-1/2}s_{\bar u}]/2}\rangle_{*,n}+o(1)\\
I_n[\beta,\zeta_h]:=&\Big(\frac{\beta}{8\pi}\Big)^{M/2}\int e^{-\beta(\bar u,\bar u)/8}d\bar u
\langle e^{\beta\mathcal{N}_n[c_1\ell_1+c_2\ell_2+n^{-1/2}s_{\bar u}]/2}\rangle_{*,n},
\end{align*}
where  $\ell_1(\lambda)=\lambda$, $\ell_2(\lambda)=\lambda^2/2$, and
\[
s_{\bar u}(\lambda)=\sum_k u_k\sqrt{\kappa_k}\dot\varphi_{h,k}(\lambda)
+n^{-1/2}(\frac{2}{\beta}-1)\zeta^{(1)}(\lambda),\quad\tilde\Phi_k\big(\bar\lambda\big)
:=\Phi_k\big(\zeta_h(\bar\lambda),\lambda_0\big).
\]
Then, changing variables once more
\[\lambda_i\to\zeta_c (\lambda):=(1-c_2/n)^{1/2}\Big(\lambda_i-\frac{c_1}{n(1-c_2/n)}\Big),\]
 we obtain that
\begin{align*}
\frac{\langle\Phi_ke^{\beta\mathcal{N}_n[h]/2}\rangle_{*,n}}
{\langle e^{\beta\mathcal{N}_n[h]/2}\rangle_{*,n}}=&\Big(\frac{\beta}{8\pi}\Big)^{M/2}
\int e^{-\beta(\bar u,\bar u)/8}d\bar u
\langle\widehat\Phi_ke^{\beta\mathcal{N}_n[n^{-1/2}s_{\bar u,c}]/2}\rangle_{*,n}\Big/\bar I_n[\beta,\zeta_h]+o(1),\\
I_n[\beta,\zeta_h]:=&\Big(\frac{\beta}{8\pi}\Big)^{M/2}\int e^{-\beta(\bar u,\bar u)/8}d\bar u
\langle e^{\beta\mathcal{N}_n[n^{-1/2}s_{\bar u,c}]/2}\rangle_{*,n},
\end{align*}
with $\widehat\Phi_k(\lambda)=\tilde\Phi_k(\zeta_c(\lambda))$ and $s_{u,c}(\lambda)=s_{u}(\zeta_c(\lambda))$.
Represent $\mathbb{R}^M=\tilde U_1\cup \tilde U_2\cup \tilde U_3$, where
\begin{align*}
&\tilde U_1=\{\bar u:(A_{h,0}\bar u,\bar u)\le\log^{2s+2}n\},\\
&\tilde  U_2=\{\bar u:\log^{2s+2}n\le(A_{h,0}\bar u,\bar u)\le C_*n\log^{4s}n\},\\
&\tilde U_3=\{\bar u:C_*n\log^{4s}n\le(A_{h,0}\bar u,\bar u) \}\\
&(A_{h,0})_{ij}=|\kappa_i|^{1/2}|\kappa_j|^{1/2}(\varphi^{(6)}_{h,i},\varphi^{(6)}_{h,j})_{\varepsilon/2}.
\end{align*}
Note, that  Proposition \ref{p:L_+} and the assumptions of Lemma \ref{l:re_h} yield
\begin{align}\label{A_0,h}
\sum(A_{h,0})_{ii}\le C\log^{2s}n,\quad ||A_{h,0}||\le C\log^{2s}n.
\end{align}
Repeating for $\tilde U_3$ the argument used for $U_5$ in Lemma \ref{l:conv}, we get
\begin{align*}
&||n^{-1/2}s_{u,c}||_\infty\le C_1n^{-1/2}(A_{h,0}\bar u,\bar u)^{1/2}\le C_2n^{-1/2}||A_{h,0}||\,||\bar u||
\le C_3n^{-1/2}\log^{2s}n||\bar u||.
\end{align*}
Hence
\begin{align*}
\Big|\langle\Phi_ke^{\beta\mathcal{N}_n[n^{-1/2}s_{u,c}]/2}\rangle_{*,n}\Big|\le (Cn)^ke^{\beta n||n^{-1/2}s_{u,c}||_\infty/2}\le
e^{c_1n^{1/2}\log^{2s}n||\bar u||}.
\end{align*}
Then, since $||\bar u||\ge(A_{h,0}\bar u,\bar u)^{1/2}/||A_{h,0}||^{1/2}\ge C_*n\log^{3s}n$ in $\tilde U_3$, we have
\begin{align*}
& \int_{\tilde U_3}d\bar u
e^{-\beta(\bar u,\bar u)/8}\Big|\langle\Phi_ke^{\beta\mathcal{N}_n[h]/2}\rangle_{*,n}\Big|\\&\le
\int_{||\bar u||^2\ge C_*n\log^{3s}n}d\bar u
e^{-\beta(\bar u,\bar u)/8+c_1n^{1/2}\log^{2s}n||\bar u||}\le e^{-nc}.
\end{align*}
For $\bar u\in\tilde U_2$
\[
||n^{-1/2}s_{u,c}^{(6)}\,||_2\le C_*\log^{2s}n,
\]
hence we can use here (\ref{Joh}) and then
 the argument used for $U_3$, but replacing $\tau$ in the Chebyshev inequality by $\tau\log^{-2s}n$.
Since by (\ref{A_0,h}) the matrix $\tau\log^{-2s}n A_{0,h}$ for small $\tau$ satisfy conditions of Proposition \ref{r:int},
we obtain that the integral in $U_2$ is $O(e^{-c\log^2n})$. Thus it suffices to study $\bar u\in\tilde U_1$. But here
\begin{align*}
&\Big|\langle\widehat\Phi_k\big(\bar\lambda\big)e^{\beta(\mathcal{N}_n[n^{-1/2}s_{\bar u}]/2}\rangle_{*,n}-
\langle\widehat\Phi_k\big(\bar\lambda\big)\rangle_{*,n}\Big|
\le\langle\widehat\Phi_k^2(\lambda)\rangle_{*,n}^{1/2}
\langle\big| e^{n^{-1/2}\mathcal{N}[s_{\bar u,c}]}-1|^2\rangle_{*,n}^{1/2},\\
&\le C n^{-1/2}((D_\sigma\Re s_{\bar u,c},\Re s_{\bar u,c})+(D_\sigma\Im s_{\bar u,c},\Im s_{\bar u,c}))^{1/2}\le n^{-\kappa}.
\end{align*}
Here we have used (\ref{VV1}) which gives the bound for  $\langle\widehat\Phi_k^2\rangle_{*,n}$
 and  (\ref{Joh}) in the case (ii), according to which  for $\bar u\in \tilde U_1$ we have for any bounded $t$
\[
\langle e^{tn^{-1/2}\mathcal{N}[s_{\bar u}]}\rangle_{*,n}=e^{t^2n^{-1}(D_\sigma s_{\bar u},s_{\bar u})}(1+o(1)).
\]
Since all $\phi_j$ of $\Phi_k$ are smooth and have  finite supports, (\ref{VV1}) imply that
\begin{align*}
\langle\widehat\Phi_k (\bar\lambda;\lambda_0)\rangle_{*,n}=&
\langle\Phi_k (\bar\lambda-n^{-1}(c_1+c_2\lambda_0/2-\tilde\zeta_h(\lambda_0))\rangle_{*,n}+O(n^{-\kappa})\\=&
\langle\Phi_k (\bar\lambda)\rangle_{*,n}+O(\varepsilon_n)+O(n^{-\kappa}).
\end{align*}
 Combining the above bounds, we get
the assertion of Lemma \ref{l:re_h}.

$\square$

\textit{Proof of Lemma \ref{l:3}.}
Introduce the
analytic function
\[f_n(t) := C_1^{-1}e^{-t^2/2}\varepsilon_n^{-1} F_n(t),\quad  t \in D.\]
 Then
\[|f_n(t)|\le 1,\quad t\in \gamma=[-\log^{1/2} \varepsilon_n^{-1}, \log^{1/2}\varepsilon_n^{-1}].\]
 Moreover, $|f_n(t)|\le C_2C_1^{-1}\varepsilon_n^{-2}$, $t\in D$.
 Then, by the theorem on two constants (see \cite{Evgr}), we conclude that
\[\log |f_n(t)|\le 2(1-\omega(t;\gamma,D))(\log\varepsilon_n^{-1}+\log(C_2/C_1)^{1/2}),\]
where $\omega(t;\gamma,D)$ is the harmonic measure of the set $\gamma$
 with respect
to the domain $D$ at the point $t\in D$. It is
well-known (see again \cite{Evgr}) that
\[\omega(t;\gamma,D)=1-\frac{2}{\pi}\Im\log
\frac{1+t/\log^{1/2}\varepsilon_n^{-1}}{1-t/\log^{1/2}\varepsilon_n^{-1}}.\]
Hence
 \[
1-\omega(t;\gamma,D)\le \frac{6\Im t}{\pi \log^{1/2}\varepsilon_n^{-1}}\le \frac{1}{4},\quad
 t\in \frac{\pi}{24} D=D',
\]
and  the above inequalities yield
\[\log |f_n(t)|\le \frac{1}{2}(\log\varepsilon_n^{-1}+\log(C_2/C_1)^{1/2})\quad\Rightarrow\quad
|f_n(t)|\le (C_2/C_1)^{1/4}\varepsilon_n^{-1/2},
\,\,\,t\in D'.\]
Then from the definition of $f_n$ we obtain (\ref{l3.2}).

$\square$
\section{Proof of Theorem \ref{t:2}}
Examining the proof of Theorem \ref{t:1}, one can  see that its result can be reformulated as follows.
For any real analytic $n$-independent one-cut $V$ and real analytic $h:||h'||,||h^{(6)}||\le \log^sn$
the inequalities hold uniformly in $h$:
\begin{align}\label{VV3}
&\Big|\frac{\langle\Phi_k(\bar\lambda,\lambda_0)e^{\beta\mathcal{N}_n[h]/2}\rangle_{V,n}}
{\langle e^{\beta\mathcal{N}_n[h]/2}\rangle_{V,n}}\Big|\le \tilde C_\Phi,\\
&\Big|\frac{\langle\Phi_k(\bar\lambda,\lambda_0)e^{\beta\mathcal{N}_n[h]/2}\rangle_{V,n}}
{\langle e^{\beta\mathcal{N}_n[h]/2}\rangle_{V,n}}-\langle\Phi_k(\bar\lambda,0)\rangle_{*,n}\Big|\le\tilde\varepsilon_n\to 0.
\label{VV4}\end{align}
Note that $\tilde C_\Phi$ and  $\tilde\varepsilon_n$ depend   on $C_\Phi$ and  $\varepsilon_n$ of (\ref{VV1}), (\ref{VV2}).

The proof of Theorem \ref{t:2} is based on these two inequalities and on the results of \cite{S:13}.
Set
\begin{align}\label{mu^*}
    \mu_\alpha=&\int_{\sigma_\alpha}\rho_\alpha(\lambda)d\lambda,\quad \rho_\alpha:=\mathbf{1}_{\sigma_\alpha}\rho,\\
V_\alpha(\lambda)=&\mathbf{1}_{\sigma_{\alpha,\varepsilon}}(\lambda)\mu_\alpha^{-1}
\Big(V(\lambda)-2\int_{\sigma\setminus\sigma_\alpha}\log|\lambda-\mu|\rho(\mu)d\mu\Big)\label{V_alpha}\\
&    \bar n:=(n_1,\dots,n_q),\quad |\bar n|:=\sum_{\alpha=1}^q n_\alpha.
\end{align}
It is easy to see that the potential $V_\alpha$ and the equilibrium density $\mu_\alpha^{-1}\rho_\alpha$
satisfy (\ref{cond_rho}) and (\ref{rho}) in $\sigma_{\alpha,\varepsilon}$,
hence $V_\alpha$ is a real analytic potential of generic
behavior in $\sigma_{\alpha,\varepsilon}$ and so in each interval we can apply  (\ref{Joh}) and (\ref{VV3})-(\ref{VV4}).

Assuming that $\lambda_0\in[a_1+\varepsilon,b_1-\varepsilon]$ and
repeating the argument of Theorem 2 of \cite{S:13}, we obtain (cf(\ref{int_Phi}))
\begin{align}\label{t2.1}
\langle\Phi_k(\bar\lambda_1;\lambda_0) \rangle_{V,n}=&\mathcal{I}^{-1}_n[V]
\sum_{|\bar n|=n}\mathcal{\kappa}_{\bar n}\Big(\frac{\beta}{2\pi}\Big)^{Mq}
\int du e^{-\frac{\beta}{8}( u,u)}\\
&\cdot\langle\Phi_k(\bar\lambda_1;\lambda_0) e^{\beta\mathcal{N}_{n_1}[
\widetilde h_1]/2}\rangle_{V_1,n_1}
\prod_{\alpha=2}^q\langle e^{\beta\mathcal{N}_{n_\alpha}[
\widetilde h_\alpha]/2}\rangle_{V_\alpha,n_\alpha},\notag\\
\mathcal{I}_n[V]=&\sum_{|\bar n|=n}\kappa_{\bar n}\Big(\frac{\beta}{2\pi}\Big)^{Mq}
\int du e^{-\frac{\beta}{8}( u,u)}
\prod_{\alpha=1}^q\langle e^{\beta\mathcal{N}_{n_\alpha}[
\widetilde h_\alpha]/2}\rangle_{V_\alpha,n_\alpha},
\notag\end{align}
where $M=[\log^2n]$, $u:=(u^{(1)},u^{(2)})$, $\kappa_{\bar n}$ are some numbers,
\begin{align}\label{t2.2}
\widetilde h_\alpha(\lambda)=&(n_\alpha-n\mu_\alpha)V_\alpha
+\dot s^{(\alpha)}(u,\lambda),
\\s^{(\alpha)}(u,\lambda)=&\sum_{j,k,\alpha'}\Big(
\widehat S_{j,\alpha';k,\alpha}u_{j,\alpha'}^{(1)}
+iS_{j,\alpha';k,\alpha}u_{j,\alpha'}^{(2)}\Big)p^{(\alpha)}_{k}(\lambda),
\notag\\
\dot s^{(\alpha)}(u,\lambda)=&s^{(\alpha)}(u,\lambda)-\frac{n}{n_\alpha}\big(s^{(\alpha)}(u,.),\rho_\alpha).
\notag
\end{align}
Thus in each interval $\sigma_{\alpha,\varepsilon}$ we are again in the situation of the one-cut analytic potential
with a "small" perturbation $\widetilde h_\alpha$. Here $\{p^{(\alpha)}_{k}\}_{k=0}^M$ are
polynomials on $\sigma_{\varepsilon}$ of degree at most $M$ and  therefore in (\ref{Joh}) we can use
the bound, valid for any $l=1,\dots,6$ and for any $\bar u$
\begin{align}\label{b_der}
(|\partial_\lambda^{(l)}s_u|,|\partial_\lambda^{(l)}s_u|)\le (CM)^{4l}\sum_\alpha\Big((D_\alpha \Re s_u,\Re s_u)+
(D_\alpha \Im s_u,\Im s_u)\Big).
\end{align}
The bound follows from the inequality, valid for polynomials  with degree not exceeding $M$:
\[\sup_{\mathrm{deg}p\le M}\dfrac{(p',p')}{(p,p)}\le CM^4,\]
which can be checked by expanding of an arbitrary polynomial
in the sum of the Jacobi polynomials orthonormal on $\sigma_\varepsilon$ without any weight.

The exact forms of  positive matrices $\widehat S=\{\widehat S_{j,\alpha';k,\alpha}\}_{\substack{j,k=1,\dots,M,\\
\alpha,\alpha'=1,\dots,q}}$
 and $S=\{ S_{j,\alpha';k,\alpha}\}_{\substack{j,k=1,\dots,M,\\
\alpha,\alpha'=1,\dots,q}}$ in (\ref{t2.2})
are not important for us. It will be important  only that (\ref{pos}) is true.

Moreover,  Lemma 2 of \cite{S:13} implies
\begin{align}\label{l:b_t}
\mathcal{T}_{\bar n}:=\kappa_{\bar n}\Big(\frac{\beta}{2\pi}\Big)^{Mq}
\int du e^{-\frac{\beta}{8}( u,u)}
\prod_{\alpha=1}^q\langle e^{\beta\mathcal{N}_{n_\alpha}[
\widetilde h_\alpha]/2}\rangle_{V_\alpha,n_\alpha}
\le Ce^{-c(\Delta n,\Delta n)},
\end{align}
where $\Delta n=(\Delta n_1,\dots,\Delta n_q)$, $\Delta n_\alpha=n_\alpha-\mu_\alpha n$, and
$\mu_\alpha$ were defined in (\ref{mu^*}). This relation and (\ref{VV3}) yield that  for our purposes it suffices
to consider in (\ref{t2.1}) only those terms for which
\begin{align}\label{b_n}
(\Delta n,\Delta n)\le c_*\log \tilde\varepsilon_n^{-1}
\end{align}
with any $n$-independent $c_*$, hence the $u$-independent part of $\widetilde h_\alpha$ of (\ref{t2.2}) cannot be too big.

Let us again  use   (\ref{Joh}) for
$\langle e^{\beta\mathcal{N}_{n_\alpha}[\widetilde h_\alpha]/2}\rangle_{V_\alpha,n_\alpha}$, $\alpha=1,\dots,q$.
Similarly to the proof of Theorem \ref{t:1}, the key
point  is that after the application of (\ref{Joh}) the real part of the correspondent quadratic form
is  negative definite, so the integral in $\bar u$ is convergent.
More precisely, Lemma 4 of \cite{S:13} guarantees that there exists $\delta>0$ such that
\begin{align}\label{pos}
\Re\Big(\sum_{\alpha=1}^q(D_{\sigma_\alpha}s^{(\alpha)},s^{(\alpha)})\Big)\le(1- \delta)(\bar u,\bar u).
\end{align}
Moreover, it is shown (see again Lemma 4, the analog of Lemma \ref{l:conv} of the present paper) that
if we define (cf (\ref{U_1}))
\begin{equation}\label{U_1.m}
U_1=\{u:=(u^{(1)},u^{(2)}):\sum_\alpha|(D_\alpha\Im \dot s_\alpha ,\Im \dot s_\alpha)|\le k_*\log n
\wedge (u^{(1)},u^{(1)})\le\log^4n\},
\end{equation}
then the integral over the complement in the r.h.s. of (\ref{t2.1}) is small.  But
if $\log\tilde\varepsilon_n^{-1}<<\log n$, the domain $U_1$ can be too big, because similarly to the proof of
Theorem \ref{t:1}, we need to consider
\begin{align*}U_0=\{u:=(u^{(1)},u^{(2)}):\sum_\alpha(D_\alpha \Re  s_\alpha , \Re s_\alpha)+
(D_\alpha \Im  s_\alpha , \Im s_\alpha)\le k_*\log \varepsilon_n^{-1}\},
\end{align*}
To estimate the integral in $U_1\setminus U_0$, one should use the Chebyshev inequality  like in (\ref{U_2}).
We are left to prove the analog of (\ref{VV4}) in $U_0$. For real $s_u$ the bound is known because of (\ref{VV4}),
and for the complex $s$ we obtain the bound from
 Lemma \ref{l:3}, repeating literally the argument used at the end of Section 2.

$\square$


\small
\begin{thebibliography}{99}

\bibitem{APS:01} Albeverio, S., Pastur, L., Shcherbina, M.: On the $1/n$
expansion for some unitary invariant ensembles of random matrices.
 Commun. Math. Phys. {\bf 224},  271-305 (2001)

\bibitem{BG:11} Borot, G. and  Guionnet, A.:
Asymptotic expansion of $\beta$-matrix models in the one-cut
regime. Comm. Math. Phys.{\bf 317}, 447-483 (2013)


\bibitem{BPS:95} Boutet de Monvel, A., Pastur L., Shcherbina M.: On the
statistical mechanics approach in the random matrix theory.
Integrated density of states.  J. Stat. Phys. \textbf{79},
585-611 (1995)

\bibitem{BI:03}
 Bleher, P., Its, A.:  Double scaling limit in the random matrix model:
the Riemann-Hilbert approach. Comm. Pure Appl. Math. {\bf 56},
433--516 (2003)

\bibitem{BEY:11} Bourgade, P., Erd\"{o}s, L., Yau, H.-T.: Universality of General $\beta$-Ensembles.\\
(http://arxiv.org/abs/1104.2272)

\bibitem{BEY:13} Bourgade, P., Erd\"{o}s, L., Yau, H.-T.: Edge Universality of Beta Ensembles
(http://arxiv.org/abs/1306.5728v1)

\bibitem{C-K:06} Claeys, T., Kuijalaars, A.B.J.: Universality of the double scaling limit
in random matrix models. Comm. Pure Appl. Math. \textbf{59}, 1573-1603 (2006)


\bibitem{DKMVZ:99} Deift, P., Kriecherbauer, T., McLaughlin, K., Venakides,
S., Zhou, X.: Uniform asymptotics for polynomials orthogonal with
respect to varying exponential weights and applications to
universality questions in random matrix theory. Commun. Pure
Appl. Math. \textbf{52}, 1335-1425 (1999)


\bibitem{De-G:07} Deift, P., Gioev, D.: Universality in random matrix
theory for orthogonal and symplectic ensembles.
Int. Math. Res. Papers. 2007, 004-116

\bibitem{De-G:07a} Deift, P., Gioev, D.: Universality at the edge of the spectrum
for unitary, orthogonal, and symplectic ensembles of random matrices.
Comm. Pure Appl. Math. \textbf{60},  867-910 (2007)

\bibitem{DGKV:07} Deift, P., Gioev, D., Kriecherbauer, T., Vanlessen, M.:
Universality for orthogonal and symplectic Laguerre-type ensembles.
J.Stat.Phys \textbf{129}, 949-1053 (2007)

\bibitem{EY:13} Erdos, L., Yau, H.-T. Gap Universality of Generalized Wigner and $\beta$-Ensembles
(http://arxiv.org/abs/1211.3786v2)

\bibitem{Evgr} Evgrafov, M.A.: Analytic Functions. Dover Pubns (1978)


\bibitem{GK:69} Gohberg, I.,  Krein, M.G.: Introduction to the Theory of Linear Nonselfadjoint Operators
American Mathematical Soc., 1969 - 378p

\bibitem{Jo:98} Johansson, K.: On fluctuations of eigenvalues of random
Hermitian matrices. Duke Math. J. \textbf{91}, 151-204 (1998)


\bibitem{L-Lub:08} Levin L., Lubinskky D.S.: Universality limits in the bulk for varying
measures. Adv. Math. \textbf{219}, 743-779 (2008)

\bibitem{Me:91} M.L.Mehta, M.L.: Random Matrices. New York: Academic
Press, (1991)

\bibitem{PS:97} Pastur, L., Shcherbina, M.: Universality of the local
eigenvalue statistics for a class of unitary invariant random
matrix ensembles. J. Stat. Phys. \textbf{86}, 109-147
(1997)

\bibitem{KS:10} Kriecherbauer,T., Shcherbina, M.:
Fluctuations of eigenvalues of matrix models and their applications.
preprint arxive: math-ph/1003.6121


\bibitem{PS:07} Pastur, L., Shcherbina, M.:
Bulk universality and related properties of Hermitian matrix models.
J.Stat.Phys. \textbf{130},  205-250 (2007)

\bibitem{PS:11} Pastur, L., Shcherbina, M.:
Eigenvalue Distribution of Large Random Matrices. Mathematical Survives and Monographs,
V171, American Mathematical Society: Providence, Rhode Island (2011)

\bibitem{RRV:07} Ram\`{i}rez, J., Rider, B., and  Vir\`{a}g, B.: Beta ensembles, stochastic Airy spectrum, and a diffusion,
J. Amer. Math. Soc. \textbf{24},  919–944 (2011)

\bibitem{S:08} Shcherbina, M.: On  Universality  for Orthogonal Ensembles of Random Matrices
Commun.Math.Phys. \textbf{285}, 957-974 (2009)


\bibitem{S:09} Shcherbina, M.: Edge  Universality  for Orthogonal Ensembles of Random Matrices
J.Stat.Phys \textbf{136}, 35-50 (2009)

\bibitem{S:11} Shcherbina, M.: Orthogonal and symplectic matrix models: universality and other properties
Commun.Math.Phys. \textbf{307},  761-790, (2011)

\bibitem{S:13} M.Shcherbina. Fluctuations of linear eigenvalue statistics of $\beta$ matrix models in the multi-cut regime
   J.Stat.Phys., \textbf{151}, N 6 ,  1004-1034 (2013)


\bibitem{Tr-Wi:98} Tracy,  C.A., Widom, H.: Correlation
functions, cluster functions, and spacing distributions for random
matrices. J.Stat.Phys. \textbf{92}, 809-835 (1998)


\bibitem{VV:09} Valk\`{o}, B., Vir\`{a}g, B.: Continuum limits of random matrices and the Brownian carousel. Invent.
Math. 177 (2009), no. 3, 463–508.
\end{thebibliography}
\end{document}